# Ultrasound in Locomotion Research - The Quest for Wider Views

**Christoph Leitner**[1,2], Christian Baumgartner[2], Christian Peham[3], Markus Tilp[1]
[1]Institute of Sport Science, University of Graz, Austria
[2]Institute of Health Care Engineering with European Testing Center of Medical Devices, Graz University of Technology, Austria
[3]Department for Companion Animals and Horses, University of Veterinary Medicine Vienna, Austria

Email: christoph.leitner@uni-graz.at

## Introduction

In a systematic review, we investigate current applications of ultrasound (US) in locomotion research (**Figure 1**). We discuss shortcomings in the range of view of US imagers and how these affect simulations of human locomotion. Furthermore, we give an outlook on emerging US technologies and approaches that can be leveraged to improve the range of view of US imagers in biomechanical applications.

To investigate locomotion and to parameterize simulation models, we need a complex laboratory setup. A combination of camera-based 3D motion capture (kinematics), 3D force plates (kinetics) and electromyographic (EMG) recordings monitor neuromuscular functions during locomotion. These methods to record human and animal locomotion patterns from an external (camera-angle) perspective have been explored and developed extensively in recent years [1-4].

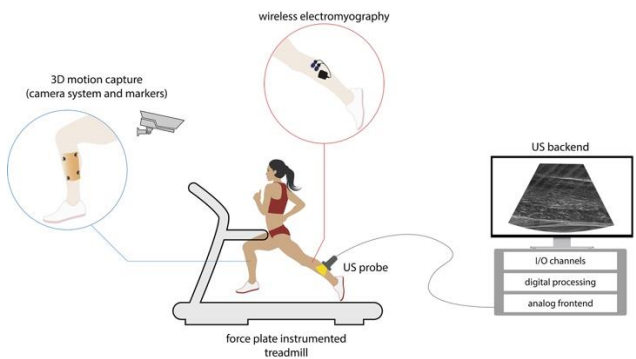

**Figure 1**: Locomotion Research Laboratory Setup [17].

However, views on internal processes (e.g. muscle and tendon dynamics) are limited or hidden to researchers. Thus, to simulate a movement and, e.g., predict forces or velocities in muscles and tendons, inverse dynamic approaches are applied:

The camera recorded gait data and the measured ground contact forces are transferred into a virtual environment using a musculoskeletal model to calculate internal parameters (e.g. muscle forces, velocities, length changes,...) for every time increment.

However, the respective contributions of muscles and tendons to the behaviour of the whole muscle and tendon unit (MTU) during locomotion are versatile. If, for example, a concentric contraction is initiated, the activated muscle shortens and aponeuroses as well as tendon(s) change in length. Thereby, the net output of an activated MTU depends on the force-velocity relation [5], the force-length relation [6], the muscle-tendon length [7], the contraction mode (e.g. eccentric, concentric, isometric contraction [8] plus contraction history effects (e.g. force enhancement and depression [9], fatigue [10], tendon-hysteresis effects [11]). It is important to understand that the interactions between muscles and tendons are responsible for the resulting movement where also storage and release of elastic energy are key [12-14].

As direct views on the behaviour of full muscle and tendon complexes during locomotion are limited, musculoskeletal simulations cannot be validated directly in-vivo and are prone to errors [15]. This has also been discussed controversially [16]. Hence, studying the muscle and tendon behaviour during locomotion in-vivo and in-silico requires a tissue measurement device with a wide range of view, a high spatio-temporal resolution and contrast [17].

## Methods

We applied the PRISMA guidelines [18] (**Figure 2**) for a systematic database research in Scopus, Medline and Google Scholar. The search code identified studies using ultrasound imaging of muscles and tendons, during locomotion, at locomotion speeds > 1.9ms$^{-1}$, in healthy subjects, and written in English language. We identified 172 studies and added 4 additional records manually. After article screening and eligibility proof 17 studies remained (**Table 1**).

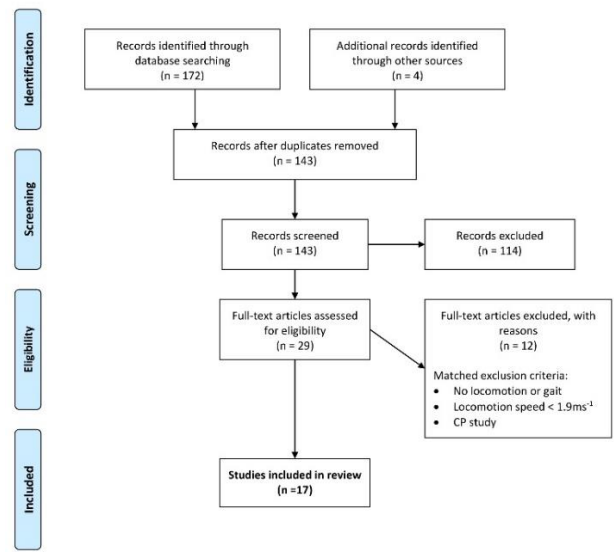

**Figure 2**: PRISMA Flow Diagram [18] of the present review.

The experimental data was extracted manually and digitally (FIGURE DIGITALIZER V1.0, Hongxue Cai, Mathworks, MA, USA) by independent surveyors.

## Results and Discussion

With the exception of Bohm et al. [21], all the studies considered in this review investigated functions of muscle fascicles in plantarflexors of the lower limbs. One reason is that plantarflexors are main contributors to human locomotion [36-38]. Furthermore, muscle fascicle lengths in the soleus (SO), medial gastrocnemius (MG), lateral gastrocnemius (LG) and tibialis posterior (TP) rarely exceed the size of the transducer at any contraction mode.

However, whole MTUs in the human plantarflexors cover larger areas [39-41] than a commercial US imager can currently capture.

As presented in **Table 1** locomotion researchers use 40 - 100 mm linear array US transducers. Thus, US systems can either locally record the muscle fascicle or the muscle-tendon junction (MTJ) movement. Therefore, a 3d motion capture system is needed to establish the relation between the locally imaged movement of landmarks (e.g., fascicle, MTJ) and the positions of body joints to which muscles connect via tendons. By combining the information of these two measurements, the MTU length can then be estimated. Some researchers applied the method proposed by Hawkins et al. [43] which provides a regression equation that relates muscle-tendon length to joint angles. Others transferred the measured data into a virtual environment (e.g., OpenSim [44,45]) where a musculoskeletal model with mainly cadaver-based data is scaled to the body anthropometry of the test subject [46-50] to calculate the length of the MTU for every time increment. Earlier studies by Lichtwark et al. [32, 35] used the estimation method proposed by Grieve et al. [51] to calculate MG MTU lengths also based on cadaver data. However, there is common consensus in favour of the methodology proposed by Fukunaga et al. [52] to estimate serial-elastic-element (SEE) lengths in the reviewed articles.

Current US developments are driven by the need to extend the range of view in 2- and 3-dimensional space. Clinical applications, for example, demand to build mobile transducers to compute 3D real time images for diagnosis (e.g., to reduce operator dependencies in examinations). Regardless of the dimension, an increase in imaging channels causes a trade-off for the computational workload. One main issue is the huge size of raw image data (>100 MB) and the amount of data rates (>10 GB/s) that need to be processed in real time. Boni et al. [53] highlight future hardware trends in extended numbers of channel system designs. In terms of biomechanical applications, US research systems might see hybrid computational approaches [54] as well as system design partitioning [55] to extend the range of view of currently used systems. Furthermore, Leitner et al. [17] have defined the frame rate as a key future engineering task and provided a list of recommendations for a new ultrasound sensors system class targeting movement science.

## Conclusions

Real time ultrasound imaging enables restricted views on in-vivo muscle and tendon behaviour during locomotion. Shortcomings in the range of view of current ultrasound systems affect the direct validation of musculoskeletal simulations as inverse approaches have to be applied. We have reviewed currently used inverse methodologies in human locomotion studies. Furthermore we presented an outlook on emerging US technologies and approaches and discussed how these developments can be leveraged for biomechanics applications.

## Acknowledgments

We want to thank Prof. Benini (ETH Zurich, Universitá di Bologna), Dr. Hager (ETH Zürich), Dr. Penasso (University of Graz) and Prof. Thaller (University of Graz) for their support in the implementation of this review.

**Table 1:** US transducer specifications, locomotion speed and investigated fascicles. Note that not all studies examined provided sufficient information (e.g., only statistical values) to extract all parameters for the investigated time intervals.

| Study | Max. locomotion speed (m/s) | Fascicle | US transducer | MTU estimation method |
|---|---|---|---|---|
| *Suzuki 2019* [19] | 5 m/s<br>*TM run[1], forefoot-strike* | MG[2] | 60 mm / 50 mm linear array | Hawkins et al. [43] |
| *Lai 2018* [20] | 5 m/s<br>*TM run* | SO[3]<br>MG<br>LG[4] | 60 mm linear array | OpenSim: Arnold et al. [48] |
| *Bohm 2018* [21] | 3 m/s<br>*TM run* | VL[5] | 100 mm linear array | Lutz et al. [55], Hawkins et al. [43] |
| *Swinnen 2018* [22] | 3.88 m/s<br>*TM run, rearfoot-strike* | MG | 60 mm linear array | OpenSim: Hamner et al. [49] |
| *Maharaj 2016* [23] | 1.9 ± 0.1 m/s<br>*TM walk, barefoot* | TP[6] | - | OpenSim: Delp et al. [46] |
| *Cronin 2016* [24] | 3 - 3.83 m/s<br>*TM run* | SO<br>MG | 50 mm | Hawkins et al. [43] |
| *Sano 2015a* [25] | 3.86 m/s<br>*TM run* | MG | 40 mm / 60 mm linear array | Hawkins et al. [43] |
| *Lai 2015* [26] | 5 m/s<br>*TM run* | SO | 60 mm linear array | OpenSim: Dorn et al. [50] |
| *Sano 2015b* [27] | 2.8 ± 0.3 m/s<br>2.9 ± 0.5 m/s<br>*TM run* | MG | 60 mm linear array | Hawkins et al. [43] |
| *Cronin 2013* [28] | 2.83 ± 0.47 m/s<br>*OG run[9], barefoot* | SO<br>MG | 60 mm linear array | - |
| *Farris 2012* [29] | 3.25 m/s<br>*TM run* | MG | - | Hawkins et al. [43] |
| *Giannakou 2011* [30] | 3 m/s<br>*TM run* | MG | 42 mm linear array | - |
| *Cronin 2011* [31] | 1.9 m/s<br>*TM run* | MG | 60 mm linear array | - |

| Lichtwark 2007 [32] | 2.08 m/s<br>*TM run* | MG | 60 mm linear array | Grieve et al. [51] |
| Ishikawa 2007a [33] | 6.5 m/s<br>*TM run* | MG | 60 mm linear array | Hawkins et al. [43] |
| Ishikawa 2007b [34] | 2.74 ± 0.21 m/s<br>*OG run* | MG | 60 mm linear array | Hawkins et al. [43] |
| Lichtwark 2006 [35] | 2.77 m/s<br>*TM run, incline* | MG | 60 mm linear array | Grieve et al. [51] |

[1] TM run—treadmill run, [2] MG—medial gastrocnemius, [3] SO—soleus, [4] LG—lateral gastrocnemius, [5] VL—vastus lateralis, [6] TP—tibialis posterior, [7] OG—overground run.